\documentclass[reprint,superscriptaddress,showpacs,preprintnumbers,amsmath,amssymb,floatfix,aps,prb]{revtex4-1}

\usepackage[dvipdfmx]{graphicx}% Include figure files
\usepackage{dcolumn}% Align table columns on decimal point
\usepackage{bm}% bold math
\usepackage[dvipdfmx]{color}%bf allow colors in manuscript

\begin{document}
%%%%% Title %%%%%
\title{
Long-range order and spin-liquid states of 
polycrystalline Tb$_{2+x}$Ti$_{2-x}$O$_{7+y}$
}
%%%%%%%%%%%%%%%%%

%%%%% Authors %%%%%
\author{T. Taniguchi}%Tomohiro Taniguchi
\affiliation{Department of Physics, Tokyo Metropolitan University, Hachioji-shi, Tokyo 192-0397, Japan}

\author{H. Kadowaki}%Hiroaki Kadowaki
\affiliation{Department of Physics, Tokyo Metropolitan University, Hachioji-shi, Tokyo 192-0397, Japan}

\author{H. Takatsu}%Hiroshi Takatsu
\affiliation{Department of Physics, Tokyo Metropolitan University, Hachioji-shi, Tokyo 192-0397, Japan}

\author{B. F{\aa}k}%Bj\"{o}rn F{\aa}k
\affiliation{SPSMS, UMR-E CEA/UJF-Grenoble-1, INAC, F-38054 Grenoble Cedex 9, France}

\author{J. Ollivier}%Jacques Ollivier
\affiliation{Institute Laue Langevin, BP156, F-38042 Grenoble, France}

\author{T. Yamazaki}%Teruo Yamazaki
\affiliation{Institute for Solid State Physics, University of Tokyo, Kashiwa 277-8581, Japan}

\author{T. J. Sato}%Taku J. Sato
\affiliation{NSL, Institute for Solid State Physics, University of Tokyo, Tokai, Ibaraki 319-1106}

\author{H. Yoshizawa}%Hideki Yoshizawa
\affiliation{NSL, Institute for Solid State Physics, University of Tokyo, Tokai, Ibaraki 319-1106}

\author{Y. Shimura}%Yasuyuki Shimura
\affiliation{Institute for Solid State Physics, University of Tokyo, Kashiwa 277-8581, Japan}

\author{T. Sakakibara}%Toshiro Sakakibara
\affiliation{Institute for Solid State Physics, University of Tokyo, Kashiwa 277-8581, Japan}

\author{T. Hong}%Tao Hong
\affiliation{Quantum Condensed Matter Division, Oak Ridge National Laboratory, Oak Ridge, Tennessee 37831-6393, USA}

\author{K. Goto}%Kazuki Goto
\affiliation{Department of Physics, Tokyo Metropolitan University, Hachioji-shi, Tokyo 192-0397, Japan}

\author{L. R. Yaraskavitch}
\affiliation{Department of Physics and Astronomy and Guelph-Waterloo Physics Institute, University of Waterloo, Waterloo, ON, Canada N2L 3G1}

\author{J. B. Kycia}
\affiliation{Department of Physics and Astronomy and Guelph-Waterloo Physics Institute, University of Waterloo, Waterloo, ON, Canada N2L 3G1}

%%%%%%%%%%%%%%%%%%%
\date{\today}

%%%%% Abstract %%%%%
\begin{abstract}
Low-temperature states of polycrystalline samples of a frustrated 
pyrochlore oxide Tb$_{2+x}$Ti$_{2-x}$O$_{7+y}$ have been 
investigated by specific heat, 
magnetic susceptibility, and neutron scattering experiments. 
We have found that this system can be tuned 
by a minute change of $x$ from a spin-liquid state 
($x < x_{\text{c}}$) to a partly ordered state with 
a small antiferromagnetic ordering of the order of  $0.1 \mu_{\text{B}}$. 
Specific heat shows a sharp peak at a phase transition at 
$T_{\text{c}}= 0.5$ K for $x=0.005$. 
Magnetic excitation spectra for this sample change from 
a quasielastic to a gapped type through $T_{\text{c}}$. 
The possibility of a Jahn-Teller transition is discussed.
\end{abstract}
%%%%%%%%%%%%%%%%%%%%

\pacs{75.10.Kt, 75.40.Cx, 75.70.Tj, 78.70.Nx}
\maketitle

Magnetic systems with geometric frustration, 
a prototype of which is antiferromagnetically coupled 
Ising spins on a triangle, have been intensively studied 
experimentally and theoretically for decades \cite{Lacroix11}. 
Spin systems on networks of triangles or tetrahedra, 
such as triangular \cite{Wannier50}, kagom\'{e} \cite{Shyozi51}, 
and pyrochlore \cite{Gardner10} lattices, play major roles 
in these studies. 
Subjects that have fascinated many investigators in recent 
years are classical and quantum spin-liquid states \cite{Bramwell01,Lee08,Balents10,Yan11}, 
where conventional long-range order (LRO) is suppressed 
to very low temperatures. 
Quantum spin-liquids \cite{Lee08,Balents10} in particular have been 
challenging both theoretically and experimentally 
since the proposal of the resonating valence-bond state \cite{Anderson73}. 
The spin ice materials R$_2$Ti$_2$O$_7$ (R = Dy, Ho) are
the well-known classical examples \cite{Bramwell01}, 
while other experimental candidates found recently 
have been studied \cite{Helton07,Itou08,Fak12,Ross11,Chang12}.

Among frustrated pyrochlore oxides \cite{Gardner10}, 
Tb$_2$Ti$_2$O$_7$ has attracted much attention 
because it does not show any conventional LRO down to 50 mK 
and remains in a dynamic spin-liquid state \cite{Gardner99,Gardner03,GG11}. 
Theoretical considerations of the crystal-field (CF) states 
of Tb$^{3+}$ and exchange and dipolar interactions of 
the system \cite{Gingras00,Enjalran04,Kao03} 
showed that it should undergo a transition into 
a magnetic LRO state at about 1.8 K within a random 
phase approximation \cite{Kao03}. 
The puzzling origin of the spin-liquid state of Tb$_2$Ti$_2$O$_7$ is 
a subject of hot debate 
\cite{Yasui02,Gardner10,Molavian07,Bonville11,Petit_TTO12,Gaulin11,Takatsu12,Fennell12,Fritsch12}. 
An interesting scenario for the spin-liquid state is the theoretical 
proposal of a quantum spin-ice state \cite{Molavian07}. 
More recently, another scenario of 
a two-singlet spin-liquid state was proposed 
to explain why inelastic neutron spectra in 
a low energy range are observed despite 
the fact that Tb$^{3+}$ is a non-Kramers ion \cite{Bonville11,Petit_TTO12}.

Several experimental puzzles of Tb$_2$Ti$_2$O$_7$ 
originate from the difficulty of controlling the quality of 
single crystalline samples, resulting in strongly sample-dependent 
specific-heat anomalies at temperatures below 2~K 
\cite{Gingras00,Hamaguchi04,Chapuis09URL,Chapuis10,Yaouanc11,Takatsu12,Ross12}. 
In contrast, 
experimental results on polycrystalline samples are more 
consistent \cite{Gardner99,Gardner03,Takatsu12}. 
Among the experimental results reported to date, 
an important clue to solve the puzzles of Tb$_2$Ti$_2$O$_7$ seems to be 
a change of state at about 0.4 K suggested by 
specific heat \cite{Takatsu12}, inelastic neutron scattering \cite{Takatsu12}, 
and neutron spin echo \cite{Gardner03} measurements on polycrystalline samples. 
At this temperature, a few single-crystalline samples show a peak 
in the specific heat suggesting a phase transition \cite{Hamaguchi04,Chapuis09URL}, 
an issue that has not been pursued seriously. 
The possibility of a cooperative Jahn-Teller phase-transition 
well below 1 K was inferred many years ago from the observation of an anomalous 
temperature dependence of the elastic constants above 1 K \cite{Mamsurova86}. 
The two-singlet spin-liquid scenario of Refs.~\onlinecite{Bonville11,Petit_TTO12,Petit12} 
is based on the assumption of a tetragonal lattice distortion 
in Tb$_2$Ti$_2$O$_7$ and the closely related ordered spin-ice 
compound Tb$_2$Sn$_2$O$_7$ \cite{Mirebeau05}, 
but the accompanying lattice distortion might be too difficult to observe directly
\cite{Lummen08,Ruff07,Nakanishi11,Gaulin11,Goto12}. 
A theoretical study on pyrochlore magnets with non-Kramers 
magnetic ground doublets, applicable to Pr$^{3+}$, Tb$^{3+}$ etc., 
pointed out the possibilities of quadrupole orderings as well as 
quantum spin ice \cite{Onoda10,Onoda11}.

In the present work, we investigate the hypothesis that the non-stoichiometry 
$x$ of Tb$_{2+x}$Ti$_{2-x}$O$_{7+y}$ is a tuning parameter 
for a quantum critical point separating a LRO state from 
a spin liquid state. 
We have therefore performed specific heat, 
magnetization, and neutron scattering experiments on 
polycrystalline samples of Tb$_{2+x}$Ti$_{2-x}$O$_{7+y}$ 
with different values of $x$. 
We find that a minute change of $x$ 
brings about a systematic change of the specific heat. 
The ground state goes from LRO 
with an unknown order parameter 
for $x > x_{\text{c}}$ to a spin liquid for $x < x_{\text{c}}$. 

\begin{figure}
\begin{center}
\includegraphics[width=7.0cm,clip]{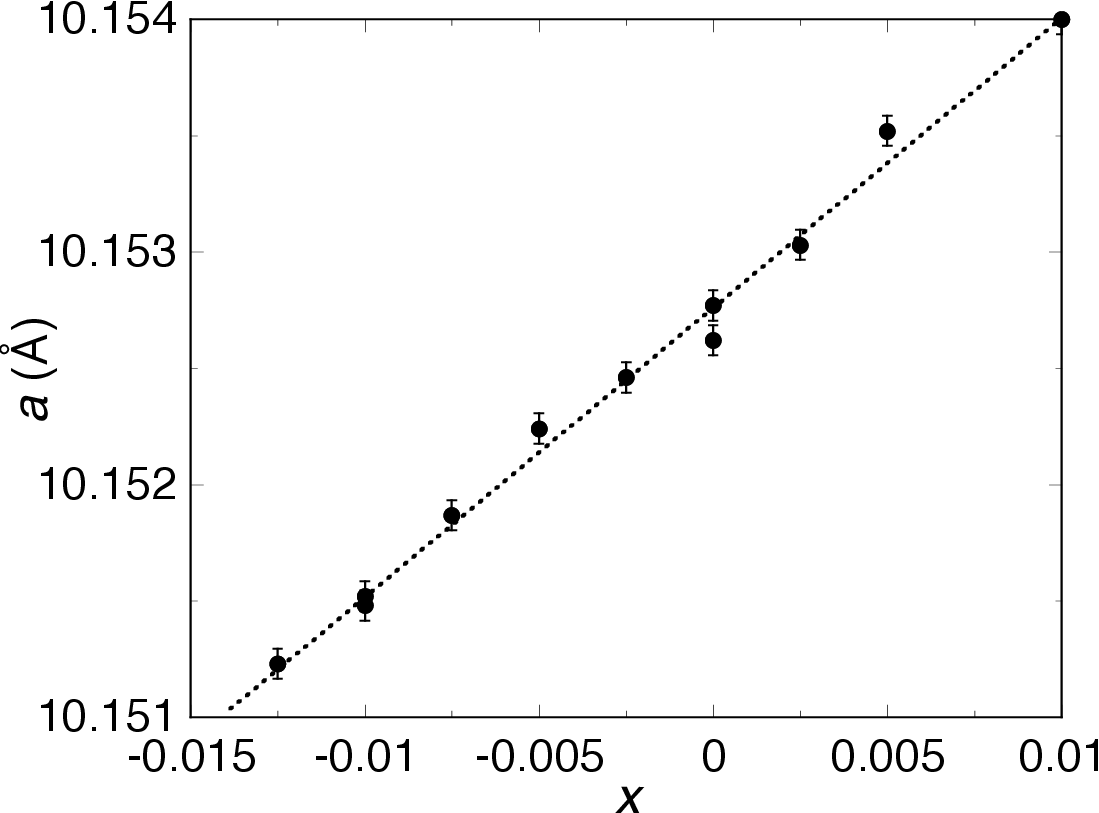}
\end{center}
\caption{
%(Color online) 
Lattice constants of polycrystalline 
Tb$_{2+x}$Ti$_{2-x}$O$_{7+y}$ at 26.0~$^{\circ}$C. 
The dashed line is a guide to the eye. 
}
\label{LC}
\end{figure}

Polycrystalline samples of Tb$_{2+x}$Ti$_{2-x}$O$_{7+y}$ with 
$-0.015 < x < 0.01$ were prepared by standard solid-state 
reaction \cite{Gardner99}. 
The value of $x$ was adjusted by changing the mass ratio 
of the two starting materials, Tb$_4$O$_{7}$ and TiO$_2$, which 
were heated in air at 1350~$^\circ$C for several days with 
periodic grindings to ensure a complete reaction. 
It was ground into powder and annealed in air 
at 800$^\circ$C for one day. 
The values of $x$ used in this paper are nominal, 
and have an offset of about $\pm 0.002$. 
The value of $y$ is determined by the oxidizing conditions. 
X-ray powder-diffraction experiments were carried out 
using a RIGAKU-SmartLab powder diffractometer 
equipped with a Cu K$_{\alpha 1}$ monochromator. 
The absence of impurity peaks in the powder diffraction patterns 
shows that the samples are single phase with the pyrochlore structure \cite{Han04}.
To measure the $x$ dependence of the lattice constant 
$a$ at 26.0~$^{\circ}$C, we performed $\theta$-$2\theta$ 
scans on powder mixtures of Tb$_{2+x}$Ti$_{2-x}$O$_{7+y}$ and Si. 
Figure~\ref{LC} shows that the lattice constant $a$, 
consistent with the previous work for $x=0$, \cite{Han04} 
has a smooth variation with $x$, which 
ensures a continuous change of the stoichiometry of 
Tb$_{2+x}$Ti$_{2-x}$O$_{7+y}$ for small $x$. 

Specific heat above 0.4 K was measured on a physical-property 
measurement-system.  
Measurements below 0.4 K were carried out using the 
quasi-adiabatic relaxation method on a dilution refrigerator \cite{Quilliam07}. 
DC magnetization measurements were carried out by a capacitive 
Faraday magnetometer in a $^3$He refrigerator. 
Neutron powder diffraction measurements were performed on 
the triple-axis spectrometer CTAX at ORNL. 
%Oak Ridge National Laboratory 
Inelastic neutron scattering measurements were carried out on 
the time-of-flight spectrometer IN5 operated 
with $\lambda = 5$ and 10 {\AA} at ILL. 
%Institute Laue-Langevin 
For these neutron scattering experiments, 
samples of $x = 0.005$ and -0.005 with weights of 5 and 9 g 
were mounted in a $^3$He (CTAX) and a dilution refrigerator 
(IN5), respectively. 

\begin{figure}
\begin{center}
\includegraphics[width=8.5cm,clip]{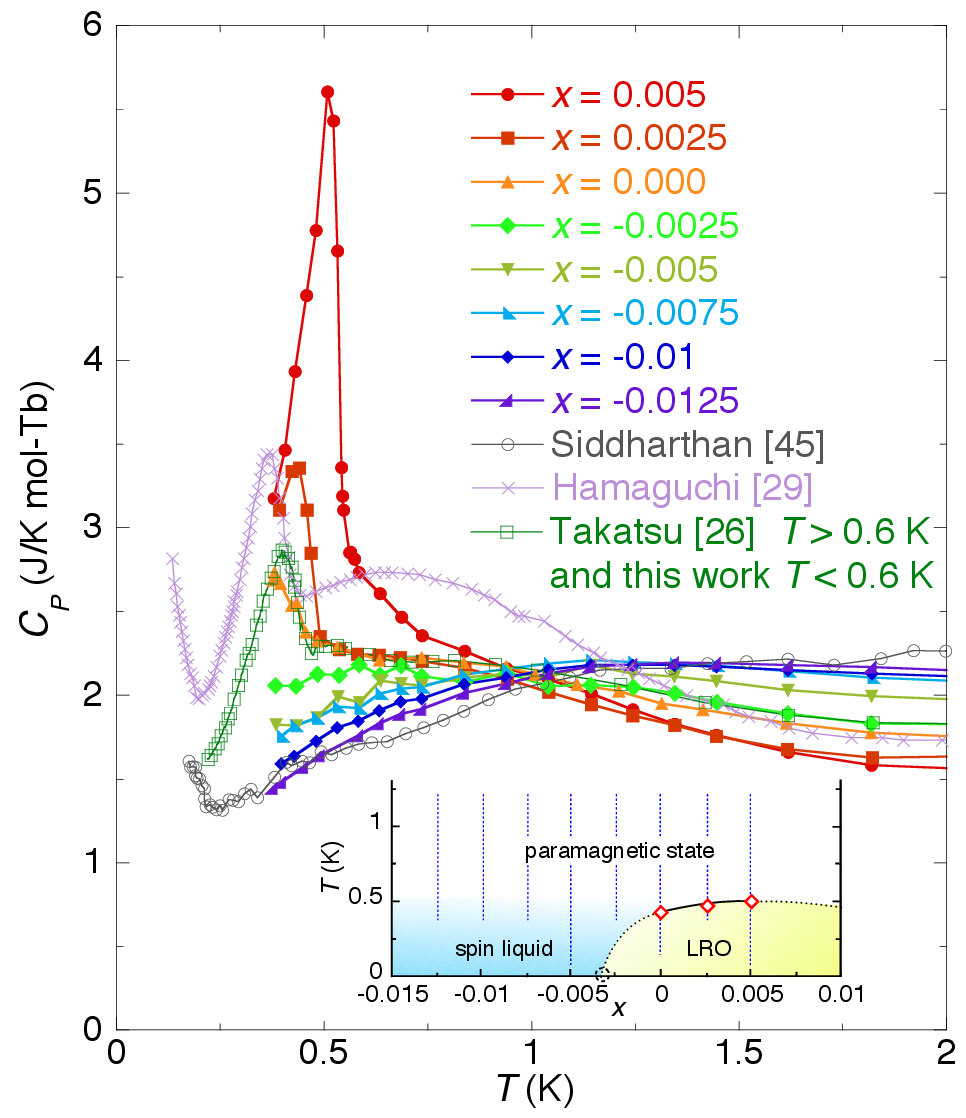}
\end{center}
\caption{
(Color online) 
Temperature dependence of the specific heat of polycrystalline 
Tb$_{2+x}$Ti$_{2-x}$O$_{7+y}$. 
Previous measurements of poly- and single-crystalline samples 
\cite{Siddharthan99,Hamaguchi04,Takatsu12}, as well as the present 
measurements below 0.6 K of a sample prepared in the same 
manner as in Ref.~\onlinecite{Takatsu12}, are plotted for comparison. 
The inset shows a phase diagram expected from the specific heat, 
susceptibility, and neutron scattering.
}
\label{Cp}
\end{figure}
In Fig.~\ref{Cp} we show the specific heat $C_P$ of the polycrystalline 
samples as a function of temperature together with a few previous 
measurements \cite{Siddharthan99,Hamaguchi04,Takatsu12}. 
Earlier work have shown qualitatively similar results \cite{Cornelius05,Ke09}. 
The $C_P(T)$ data show a systematic change by varying $x$. 
A sample with $x = 0.005$ shows a clear peak indicating 
a second-order phase transition at $T_{\text{c}} = 0.5$ K. 
Samples with $x = 0.0025$ and 0.000 show smaller peaks 
at 0.43 and 0.4 K, respectively. 
We note that $C_P$ of the present sample with $x=0.000$ 
agrees approximately with our previous measurements \cite{Takatsu12}, 
the temperature range of which was extended down to 0.2 K in the present work 
on a sample (nominal $x^{\prime} = 0$) prepared from a different 
commercial source of Tb$_4$O$_7$. 
Our previous interpretation \cite{Takatsu12} of 
the upturn below 0.5 K as a crossover behavior is incorrect 
due to the insufficient temperature range. 
The previous $C_P$ data \cite{Siddharthan99} (Fig.~\ref{Cp}) 
on a polycrystalline sample with their nominal $x^{\prime \prime} = 0$ 
corresponds to our $x = -0.0125$, implying that fine tuning of 
$x$ requires careful sample preparation. 
In the inset of Fig.~\ref{Cp}, we show a cumulative phase diagram constructed 
from $C_P(T,x)$ in conjunction with 
the susceptibility and neutron scattering experiments 
discussed below. 

A peak of $C_P(T)$ in Tb$_2$Ti$_2$O$_7$ was first reported 
for a single-crystalline sample at 0.37 K \cite{Hamaguchi04}. 
These $C_P(T)$ data \cite{Hamaguchi04}, reproduced in Fig.~\ref{Cp},  
show a significantly different $T$ dependence from any of 
the polycrystalline samples. 
The sharp peak at 0.37 K may result from a portion of 
the sample having a non-stoichiometry parameter around 
$x = -0.001$, corresponding to a peak slightly lower 
in temperature than our $x = 0.000$. 
However, a hump in $C_P(T)$ around 0.75 K for the single crystal 
does not appear for the polycrystalline samples. 
We believe that these single- and poly-crystalline samples have 
significant, but presently not well understood, differences in quality. 

\begin{figure}
\begin{center}
\includegraphics[width=7.0cm,clip]{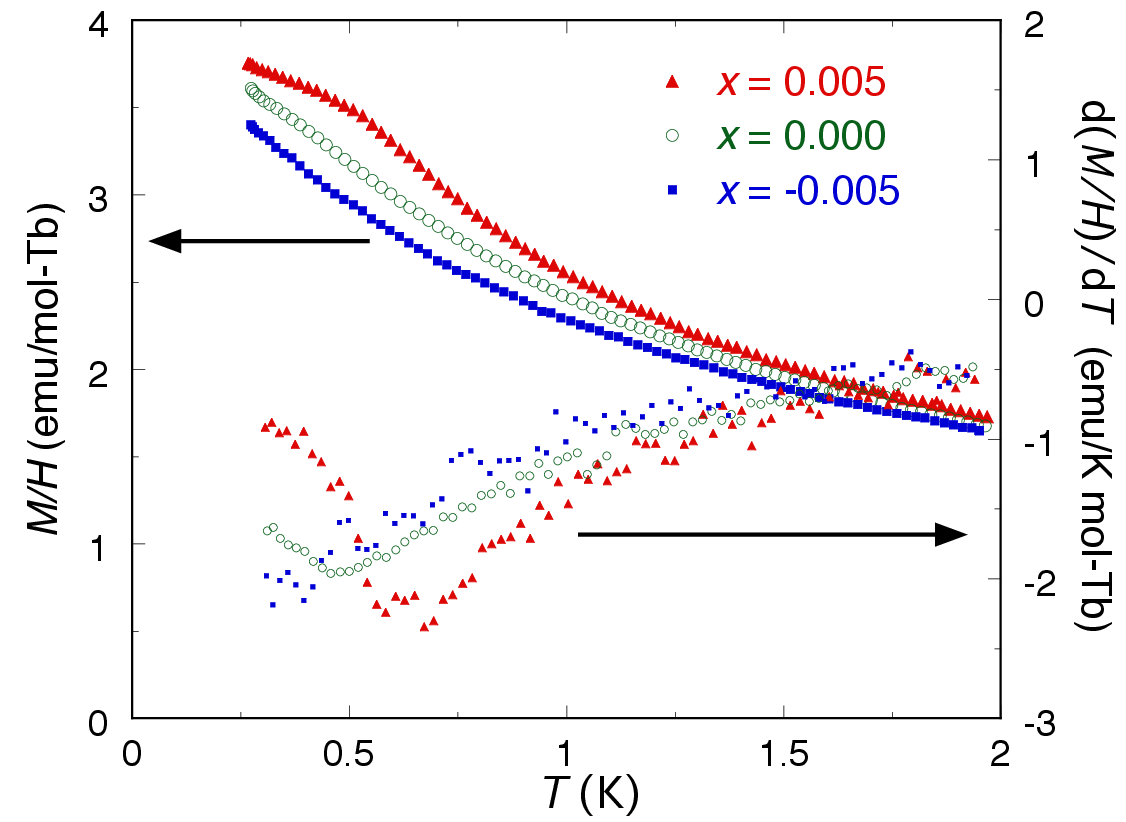}
\end{center}
\caption{
(Color online) 
Temperature dependence of the magnetic susceptibility 
($H=0.05$ T) and its derivative with respect to $T$ 
of polycrystalline Tb$_{2+x}$Ti$_{2-x}$O$_{7+y}$ with 
$x = -0.005$, 0.000, and 0.005. 
}
\label{CHI}
\end{figure}

In order to check whether $T_{\text{c}}$ is an antiferromagnetic 
transition, as suggested in Ref.~\onlinecite{Hamaguchi04}, 
we performed magnetization and neutron powder-diffraction experiments. 
In Fig.~\ref{CHI} we show the magnetic susceptibility as a function 
of temperature for three polycrystalline samples 
with $x = \pm 0.005$ and 0.000. 
The susceptibilities for $x=0.005$ and 0.000 
show only slight anomalies around the 
clear peaks of $C_P(T)$ at $T_{\text{c}}=0.5$ and 0.4 K, 
respectively. 
These weak anomalies resemble certain transitions 
related to magnetic degrees of freedom. 

\begin{figure}
\begin{center}
\includegraphics[width=8.5cm,clip]{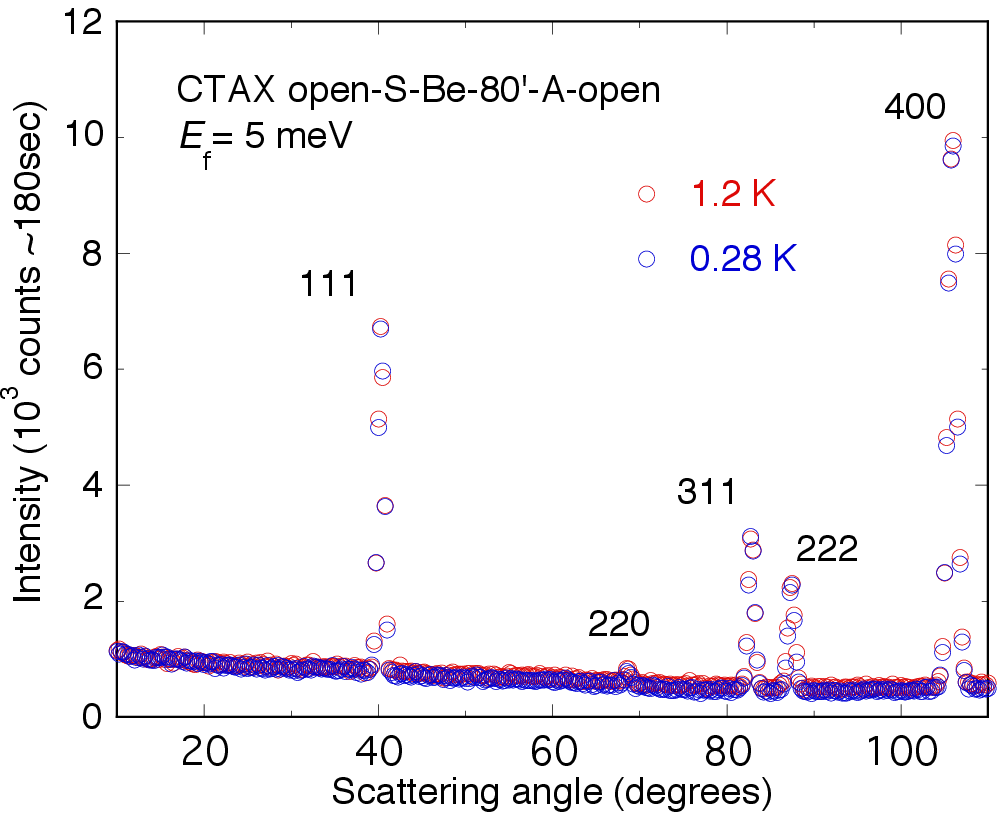}
\end{center}
\caption{
(Color online) 
Neutron powder diffraction pattern of polycrystalline 
Tb$_{2+x}$Ti$_{2-x}$O$_{7+y}$ with 
$x = 0.005$ taken above and below $T_{\text{c}} = 0.5$ K. 
}
\label{NPD}
\end{figure}

In Fig.~\ref{NPD} we show neutron powder-diffraction patterns 
for the $x=0.005$ sample below and above $T_{\text{c}}$. 
The pattern below $T_{\text{c}}$ shows 
neither any clear antiferromagnetic reflections 
nor any clear changes due to a structural transition. 
Rough estimates of the upper limits of the antiferromagnetic 
ordered moment and the structural change are about 0.1 $\mu_{\text{B}}$ 
and $\Delta a/a < 0.01$ assuming a cubic to tetragonal 
distortion. 
The intensity of the sloping paramagnetic scattering, 
a background for Bragg peaks, decreases slightly 
as temperature is lowered from 1.2 to 0.28 K. 
This is brought about by a change in the magnetic excitations. 
The lack of obvious antiferromagnetism distinctly separates 
Tb$_2$Ti$_2$O$_7$ from the ordered spin-ice compound Tb$_2$Sn$_2$O$_7$ 
\cite{Mirebeau05,Mirebeau07},
in which antiferromagnetic ordering with a moment of 5.9 $\mu_{\text B}$ 
was observed well below $T_{\text{c}}= 0.87$ K. 

\begin{figure}
\begin{center}
\includegraphics[width=8.5cm,clip]{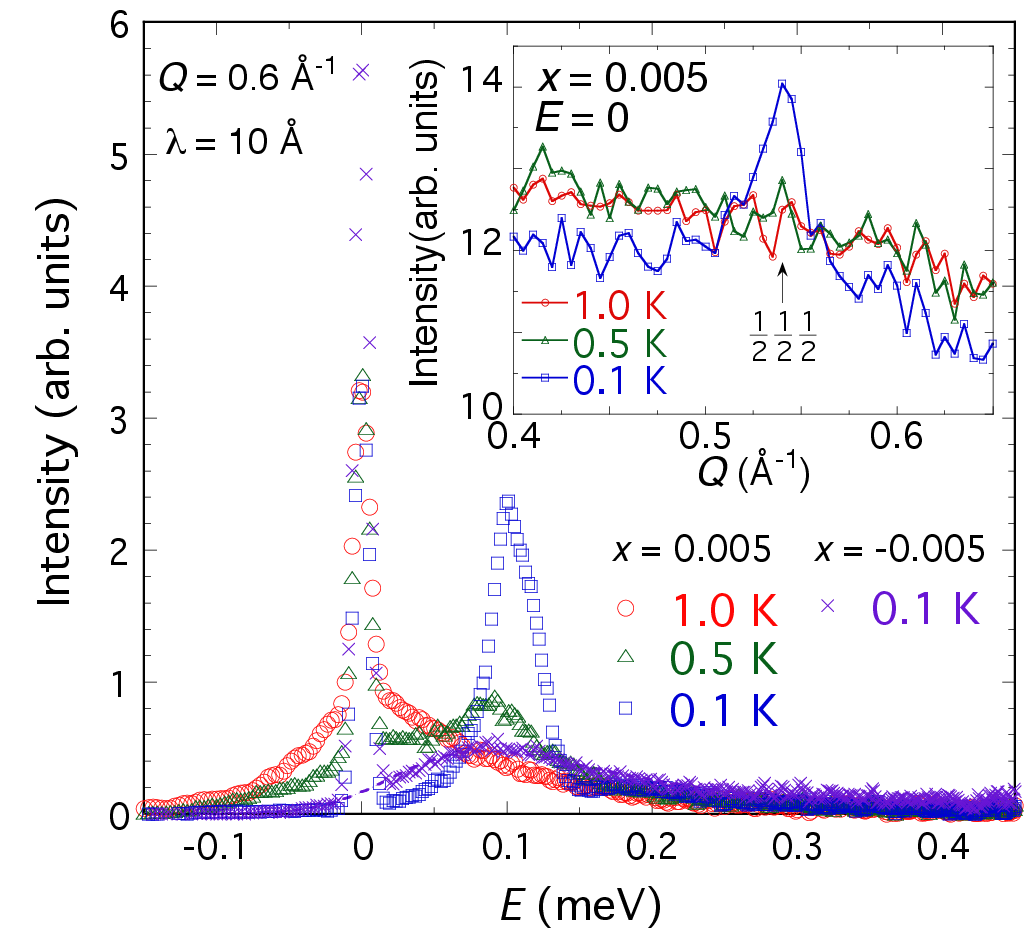}
\end{center}
\caption{
(Color online) 
Energy spectra of inelastic neutron scattering for polycrystalline 
Tb$_{2+x}$Ti$_{2-x}$O$_{7+y}$ with $x = 0.005$ and -0.005. 
The inset shows the $Q$ dependence of the elastic scattering for 
the $x = 0.005$ sample 
around $Q = |(\frac{1}{2} \frac{1}{2} \frac{1}{2})|$ 
above and below $T_{\text{c}}$. 
The dashed line is the fit curve. 
}
\label{INS}
\end{figure}

To study the spectral change of the magnetic excitations through $T_{\text{c}}$, 
we performed inelastic neutron scattering measurements using the 
spectrometer IN5 \cite{Ollivier11} with 
an energy resolution of $\Delta E = 0.012$ meV (FWHM), 
which is 5 times better than that in our previous study \cite{Takatsu12}. 
Figure~\ref{INS} shows the temperature dependence of 
an energy spectrum for the $x=0.005$ sample at $Q = 0.6$ {\AA}$^{-1}$. 
It is evident that the spectrum changes from 
a continuum ($T > T_{\text{c}}$) to a peaked structure at 
0.1 meV ($T < T_{\text{c}}$). 
The excitation peak at $T \ll T_{\text{c}}$ is weakly 
$Q$-dependent, which may possibly be interpreted 
as a splitting of the CF ground-state doublet. 
An energy spectrum of the $x=-0.005$ sample is also shown 
in Fig.~\ref{INS} for comparison. 
Its spectral shape can be approximately expressed by 
a Lorentzian squared 
$\text{Im}\chi(E,Q)/E \propto [(\sqrt{2}-1)E^2+\Gamma^2]^{-2} $ 
with $\Gamma=$ 0.1 meV (HWHM) in $-0.05 < E < 0.3$ meV, 
revealing quantum spin fluctuations with 
the same energy scale of 0.1 meV as that of the $x=0.005$ sample. 

The high sensitivity of IN5 enabled us to observe a small 
Bragg peak for the $x=0.005$ sample, being undetectable in the CTAX data (Fig.~\ref{NPD}). 
In the inset of Fig.~\ref{INS}, the intensity of the elastic scattering for 
$|E| < 0.005$ meV is plotted as a function of $Q$. 
Below $T_{\text{c}}$, a clear Bragg peak at $Q=0.54$~\AA$^{-1}$ 
is observed, which can be indexed as $(\frac{1}{2} \frac{1}{2} \frac{1}{2})$. 
The $Q$-width of this peak is somewhat larger than the instrumental 
$Q$-resolution, 
and corresponds to a correlation length 
of the order of 100 \AA. 
Although this peak could be of a nuclear (structural) origin, 
it is more likely an antiferromagnetic (AFM) reflection. 
In fact, two recent neutron scattering 
experiments carried out on single-crystalline samples of 
Tb$_{2}$Ti$_{2}$O$_{7}$ showed magnetic short-range order 
around the same $\bm{Q}=(\frac{1}{2} \frac{1}{2} \frac{1}{2})$ \cite{Fritsch12,Petit_TTO12}. 
A roughly estimated ordered moment for the $x=0.005$ sample 
is 0.08 $\mu_{\text B}$ at 0.1 K, 
where we assume the phase factor $e^{i \bm{Q} \cdot \bm{r}}=1$ 
in the magnetic structure factor. 
This ordered moment is much smaller than the magnetic moment 
$\sim$ 5 $\mu_{\text{B}}$ of the ground doublets \cite{Gingras00,Mirebeau07}, 
which implies that most of the spin fluctuations persist below $T_{\text{c}}$. 
In contrast, the entropy change around 
$T_{\text{c}} = 0.5$ K is $S(T=0.55)-S(T=0.38) \simeq 0.25 R \ln 2$ (Fig.~\ref{Cp}), 
which is significant. 
These probably indicate that there is a major order parameter, 
which is unknown at present.

The present 
results have provided an answer to the problem of 
the previously reported transition or crossover at about 0.4 K 
for the poly- and single-crystalline Tb$_{2}$Ti$_{2}$O$_{7}$ 
\cite{Gardner03,Hamaguchi04,Takatsu12}, 
and they posed another question: what is the major order parameter 
associated with $T_{\text{c}}$? 
In the following, we speculatively discuss a few possibilities. 
A cooperative Jahn-Teller transition due to a magneto-elastic 
coupling \cite{Jensen91,Gehring75} was suggested a long time ago \cite{Mamsurova86}, 
although direct experimental evidence has not been found. 
Precursor effects of a Jahn-Teller transition were reported 
using X-ray diffraction on a single-crystalline sample \cite{Ruff07}.
According to Refs.~\onlinecite{Bonville11,Petit_TTO12,Petit12}, 
a splitting of the CF ground-state doublet into two singlets 
can be interpreted as simplest evidence of a Jahn-Teller 
distortion breaking the local trigonal $D_{3d}$ symmetry of the 
Tb site. 
Along these lines, the weakly $Q$-dependent excitation peak at 0.1 meV 
(Fig.~\ref{INS}) can be interpreted as the splitting, 
and the transition is due to a Jahn-Teller effect 
accompanying a small AFM ordering \cite{Gehring75}. 
A recent theory \cite{Onoda10,Onoda11}, 
exploited to explain quantum fluctuations 
of pyrochlore magnets with non-Kramers Pr$^{3+}$, Tb$^{3+}$ etc., 
showed possibilities of quadrupole orderings due to an electronic coupling, 
which are located close to the quantum spin ice state \cite{Molavian07}. 
One of these quadrupole orderings \cite{Onoda10,Onoda11} may be the order parameter. 
A resulting structural distortion coupled to the quadrupole ordering 
could be too small to be observed. 
Although the major order parameter is unknown at present, 
the long-standing puzzle of the spin-liquid state 
of Tb$_{2}$Ti$_{2}$O$_{7}$ seems to be reformulated to a novel problem 
of frustration having spin and other degrees of freedom. 
Experimentally, single-crystalline samples with tunable 
$x$ or $y$ are indispensable for further studies. 
%}

In summary, 
we have investigated the low-temperature states of polycrystalline 
Tb$_{2+x}$Ti$_{2-x}$O$_{7+y}$ samples by specific heat, 
magnetic susceptibility, and neutron scattering experiments. 
We have found that this system can be tuned by a minute 
change of $x$ from a LRO ground state 
with an unknown major order parameter 
accompanying a minor AFM ordering 
for $x > x_{\text{c}}$ 
to a liquid-type ground state with quantum spin-fluctuations 
for $x < x_{\text{c}}$. 
Specific heat shows a sharp peak at a second-order phase-transition 
$T_{\text{c}}$ for $x > x_{\text{c}}$. 
Inelastic neutron scattering of an $x=0.005$ ($> x_{\text{c}}$) sample 
shows that a gap opens in the magnetic excitation spectrum below 
$T_{\text{c}}$.

%
%\begin{acknowledgments}
We thank M.J.P. Gingras, R. Higashinaka, J.W. Lynn, and K. Matsuhira 
for useful discussions. 
This work was supported by KAKENHI NSMIF.
The specific heat to 0.4 K and magnetization measurements 
were performed using facilities of ISSP, Univ. of Tokyo. 
Work on CTAX was supported by the 
US-Japan Cooperative Program on Neutron Scattering. 
HFIR was partially supported by the US DOE, Office of BES, 
Division of Scientific User Facilities. 
The neutron scattering performed using IN5 (France) 
was transferred from JRR3-HER (proposal 11567) 
with the approval of ISSP, Univ. of Tokyo, and JAEA, Tokai, Japan.
%\end{acknowledgments}
%

\bibliography{TTO_TT}
\end{document}